\input {epsf}
\documentstyle[prl,aps, multicol,epsf]{revtex}
\begin{document}
\draft
\title{Numerical Investigation of the Dynamics of a Thin \\ Film Type II 
  Superconductor with and without Disorder} 
\author{A.K.Kienappel and M.A.Moore}
\address{Department of Physics, University of Manchester,
Manchester, M13 9PL, United Kingdom.}
\date{\today}
\maketitle
\begin{abstract}
The equilibrium 
dynamics of a thin film type II superconductor with spherical geometry
are investigated numerically in a simulation based on the lowest Landau 
level approximation to the time-dependent Ginzburg-Landau equation.
Both the static and time-dependent density-density correlation functions 
of the superconducting order parameter have
been investigated for systems with varying amounts of
 quenched random disorder.
 As the temperature is lowered it is found that the correlation length,
the length-scale over which
the vortices have short-range crystalline order, increases but 
the introduction of quenched random disorder reduces  this
correlation length. We see no signs of a phase transition in either the 
pure or the disordered case.
For the disordered system there is no evidence for the existence 
 of a Bragg glass phase with quasi long-range correlations.
The dynamics  in both the  pure and disordered systems is activated, 
and  the barrier of the relaxation mechanism grows linearly 
with the correlation length. 
The self-diffusion time scale of the vortices 
was also measured and has 
the same temperature dependence
as that of the longest time scales
found in the time dependent density-density correlation function.  
The dominant relaxation mechanism observed is a change in orientation of a 
correlated region of size of the correlation length. 
A scaling argument is given to explain the value of the barrier exponent.  

\end{abstract}

\begin{multicols}{2}
\narrowtext

\section{INTRODUCTION}
The nature of the mixed state of thin films of type II
 superconductors is despite considerable 
experimental, theoretical and numerical research still unclear.
 In the mean-field limit, which describes the ground state
of the system,  the vortices are known 
 to form into a triangular Abrikosov lattice \cite{parks}. 
Also it is generally accepted that close to the mean-field 
transition line 
to the normal state, the $H_{c2}$ line, thermal fluctuations
destroy that hexagonal order 
and lead to a vortex liquid state \cite{eilenberger}.
However, the phase diagram between the mean-field $H_{c1}$ and
 $H_{c2}$ lines  
remains unclear. The Kosterlitz-Thouless-Halperin-Nelson (KTHN)
theory of a two dimensional (2D) 
melting \cite{halperin&nelson} has been 
applied to superconductors \cite{doniach&hubermann}
\cite{fisher} yielding a phase diagram with a solid, a hexatic and 
a liquid phase.
Other theoretical work raises a doubt as to whether there is {\it any}
 phase transition to the mixed state and suggests that a flux lattice
exists only in the zero temperature limit  and that the liquid phase 
exists at any non-zero temperature \cite{moore-odlro} \cite{Yeo}. 
It has long been predicted that in the presence of disorder 
any long range crystalline order of a vortex lattice phase is destroyed
in less than four dimensions \cite{larkin}. 
In the  glass like state which is expected to form instead, 
the length scale of the short-range crystalline order has been 
predicted by Ovchinnikov and Larkin \cite{larkin&ovchinnikov} for the
zero-temperature (mean-field) limit.
More recently the existence of a Bragg glass phase with quasi 
long range order has been suggested \cite{giamarchi&doussal}.
The non-perturbative analytical method used by Yeo and Moore \cite{Yeo},
which yields a 2D flux liquid at all temperatures,
 has also been applied to the case
of quenched random disorder, which is found to just reduce the extent  of 
short-range crystalline order present
 in the liquid state. We shall find  good agreement 
of our numerical results with this work, but none with the KTHN picture, 
or in the presence of disorder with the Bragg glass scenario.  

A lot of numerical work has been done on clean systems, 
some indicating a first order 2D vortex melting transition 
\cite{hu_macdonald} \cite{kato_nagosa} and some
failing to see any kind of phase transition \cite{o'neill} \cite{lee} 
\cite{dodgson}. 
The experimental evidence on superconducting films is also contradictory. 
Whilst a first order 2D melting transition has never been observed,   
current-voltage measurements have provided some evidence for 
second order 2D melting \cite{berghuis_kes} as well as failing to 
provide evidence for any transition in other experiments \cite{nikulov}.
Yazdani et al. \cite{yazdani} have reported  evidence for 2D melting 
from measurements of ac penetration depth in thin films of $\alpha$- MoGe. 
Theunissen et al. \cite{kes_szef} find indication of a melting as well 
as a hexatic to liquid transition in  measurements of vortex viscosity 
in NbGe. 

Most of this experimental evidence is on transport phenomena in 
samples that have at least weak disorder or surface pinning 
due to crystal defects or impurities. 
This focuses  interest on the dynamics of the mixed state and the 
influence of quenched random disorder. For the description of ac transport 
phenomena equilibrium dynamics is relevant. It is also vital for the 
whole picture of the mixed state. 
It has been suggested that the apparent second order freezing transition 
deduced 
from changes in vortex liquid viscosity or sample resistance 
may instead be due to an exponentially fast increase of the relaxation time
scales at low temperatures.\cite{moore-PhT?}.
This behavior, which our work confirms, indicates activated 
dynamics, in which the relaxational time scales increase exponentially   
with the correlation length.

In this paper we present a numerical investigation into relaxational 
time scales of density correlations in the vortex liquid and 
diffusion time scales of the vortices in a thin film.
 A Langevin dynamics simulation  is used with a phenomenological
Ginzburg-Landau Hamiltonian. To simulate quenched random disorder, we add 
a Gaussianly-distributed random contribution to the mean-field
transition temperature in the Hamiltonian.
 The equilibrium dynamics of the superconductor is investigated
by measurements of the time scales of density correlation decay and 
vortex self-diffusion. We find that the time scales of  the
two different processes both show the same exponential growth with the
translational correlation length in the system  and are 
due to the same activated relaxation process.

The paper is organized as follows. 
In Section \ref{sec:tf} we shall give a short summary of the background 
of the phenomenological Ginzburg-Landau theory 
on which our work is based. We then give a description of how we solve 
the time dependent Ginzburg-Landau equation  numerically in
Section \ref{sec:ns}. Section \ref{sec:dc} reports  our work 
on the density correlations and their relaxation times, followed by  
Section \ref{sec:sdc} which reports on self-diffusion of the vortices.
 In Section \ref{sec:ad} we put together our results
from the previous sections and interpret them. We also 
identify the underlying dynamic processes.  
In Section \ref{sec:ccl} we summarize and discuss our results.

\section{THEORETICAL FRAMEWORK}
\label{sec:tf} 
Our simulation is based on the phenomenological Ginzburg-Landau theory 
in the approximation of a uniform magnetic field  ${\bf B}$.  
The free energy functional for a clean sample is given by 
\begin{equation}
\label{eq:GL_fe}
F\!=\!\int\! d^{3}\!r \left (\alpha|\psi|^{2}+\frac{\beta}{2}|\psi|^{4}
              +\frac{1}{2m}|(-i\hbar {\bf \nabla}-2e{\bf A})\psi|^2
              \right),
\end {equation}
where $\psi$ is the order parameter representing the macroscopic 
wave function of the superconducting 
electrons and ${\bf B}={\bf \nabla \times A}$.
In first approximation $\alpha(T)=\alpha '(T-T_{c})$
and $\beta(T)$ is a constant.  $ \alpha ',\beta>0$.

For the case of quenched disorder a random local potential is added to 
the free energy:
\begin{equation}
\label{eq:dis_fe}
F_{dis}=\int d^{3}r \,\Theta({\bf r})|\psi({\bf r})|^2,
\end {equation}
where $\Theta({\bf r})$ is real and  Gaussian distributed with 
\begin{equation}
\label{eq:av_dis}
\langle\Theta({\bf r})\rangle=0,
\end{equation}
\begin{equation}
\label{eq:corr_dis}
\langle\Theta({\bf r})\Theta({\bf r'})\rangle=
               \Delta\delta({\bf r}-{\bf r'}).
\end{equation}
Angular brackets denote thermal averages and $\delta$ is the 
three dimensional Dirac delta 
function. $\Delta$ is the measure of the strength of the disorder.

The simulation follows Langevin dynamics, described by the time dependent
Ginzburg-Landau equation.
\begin{equation}
\label{eq:TDGL}
\frac{\partial\psi({\bf r},t)}{\partial t}=-\Gamma 
\       \frac{\partial F}{\partial \psi^{\ast}}+\xi({\bf r},t),  
\end{equation}
where  $F$ is defined in Eq. (\ref{eq:GL_fe}) and $\xi$ is Gaussian white
noise of strength $2\Gamma k_{B}T$, i.e. 
\[ 
\langle\xi^{\ast}({\bf r},t)\, \xi({\bf r}',t')\rangle = 
    2\Gamma\, k_{B}T\,\delta({\bf r-r}')\delta(t-t')\;.
\]

\section{NUMERICAL SOLUTION OF THE TIME DEPENDENT GINZBURG-LANDAU EQUATION}
\label{sec:ns} 

We base our simulation on a thin film superconductor 
with the following experimentally realizable  two dimensional geometry. 
A thin film superconducting sphere of radius $R$ and thickness $d$ 
is placed in a radial magnetic field of magnitude 
$B$ which emerges from an infinitely long, thin solenoid whose end
is at the center of the sphere. 
This system has been investigated numerically before, 
using Monte Carlo dynamics \cite{o'neill} \cite{lee} \cite{dodgson}. 
The reasons for our preference of this geometry to the more
widely used geometry of a plane with periodic boundary conditions 
have been presented in detail by Dodgson and Moore \cite{dodgson}. 
The main advantage is that the
spherical geometry guarantees full translational symmetry, which periodic
boundary conditions do not, while having the same thermodynamic limit.
 Periodic boundary conditions impose an artificial pinning potential 
on the vortices which makes them unsuitable for investigating transport
and dynamical phenomena.

We assume that in a strong type II superconductor where 
the magnetic penetration depth is larger than the coherence length of the
superconducting wave function by orders of magnitude, fluctuations in the
magnetic induction are negligible.
Flux quantization requires that the B field at the sphere be such that
 $4\pi\,R^{2}\,B= N\Phi_{0}$, where $\Phi_{0}=h/2e$ is the flux quantum 
and $N$ is the number of vortices.
This  fixes $R=\sqrt{N/2}$ with the unit of length being 
$l_{m}=\sqrt{\Phi_{0}/2\pi B}$, which is proportional
to the nearest neighbor distance of vortices in the Abrikosov lattice. 

We shall make the usual approximation
of retaining only  the lowest Landau level (LLL). This is the
approximation that  Abrikosov used to first describe the
vortex lattice state
at mean-field level \cite{parks}.
Defining ${\bf D}$ as the gauge invariant momentum operator, 
${\bf D}=-i\hbar{\bf \nabla} - 2e{\bf A}$, the LLL is the 
eigenspace 
of $D^{2}$ associated with its lowest eigenvalue $2eB\hbar$.  
The LLL approximation is traditionally believed to hold 
near the upper critical field $H_{c2}$. However, due to renormalization 
effects, it actually describes a large portion of the vortex liquid 
regime \cite{ikeda}.
 
In spherical geometry an orthonormal basis of the LLL is \cite{roy}
\begin{equation}
 \psi_{m}(\theta,\phi)=(4\pi R^{2})^{-1/2} A_{m}
e^{im\phi}\sin^{m}(\theta /2)\cos^{N-m}(\theta /2), 
\label{eq:lll}
\end{equation}
where $N$ is the number of vortices, $0\leq m\leq N$, and 
$A_{m}=B(m+1, N-m+1)^{-1/2}$. $B$ stands for the Beta 
function, given for natural numbers n,m by
\[B(n,m)=\frac{(n-1)!\,(m-1)!}{(n+m-1)!}.\]
The order parameter in the LLL approximation can be expanded in
the above basis set:  
\begin{equation}
\label{eq:phiexp}
{\psi(\theta,\phi)}=Q\sum_{m=0}^{N}v_{m}\psi_{m}(\theta,\phi),
\end{equation}
 where
$Q=(\Phi_{0}k_{B}T/\beta d B)^{1/4}$ \cite{dodgson}.

The free energy Hamiltonian for our system is given as a sum of the
Hamiltonian of the clean system and the contribution from disorder,
${\cal H}={\cal H}_{cl}+{\cal H}_{dis}$, where 
${\cal H}_{cl}$ and ${\cal H}_{dis}$ are the free energy terms 
from Eqs. (\ref{eq:GL_fe}) and (\ref{eq:dis_fe}) respectively.
If the order parameter is now expanded 
in the LLL eigenstates, we can carry out the spatial 
integral to express the Hamiltonian of a system without disorder 
in terms of the LLL coefficients $v_{m}$:
\begin{equation}
\label{eq:simham}
\frac{{\cal H}_{cl}(\{v_{m}\})}{k_{B}T}
=\alpha_{T}\sum_{m=0}^{N}|v_{m}|^{2}
                            +\frac{1}{2N}\sum_{p=0}^{2N}|U_{p}|^{2}.
\end{equation}
The effective temperature  parameter is given by
\begin{equation}
\alpha_{T}=\frac{dQ^{2}}{k_{B}T} \left(\alpha(T)+\frac{eB\hbar}{m}\right).
\end{equation}
Note that $\alpha_{T}=0$ corresponds to the mean-field $H_{c2}(T)$ line and
$\alpha_{T}=-\infty$ to $T=0$.
In the quartic term,
\begin{equation}
U_{p}=\sum_{q=0}^{N}\theta(p-q)\theta(N-(p-q))\,
f_{N}(q,p-q)\,v_{q}v_{p-q},
\end{equation}
where $\theta$ is the Heaviside step function and 
\[ 
f_{N}(m,n)=(A_{m}A_{n}\,B(m+n+1,\,2N-m-n+1))^{1/2}.
\]

To express the disorder contribution ${\cal H}_{dis}$ in terms of the LLL
 coefficients $v_{m}$, we expand the random Gaussian disorder on 
the sphere in normalized spherical harmonics $Y_{l}^{m}$:
\[ 
\Theta(\theta,\phi)= \frac{k_{B}T}{dQ^{2}}\sum_{l=0}^{\infty}
                        \sum_{m=-l}^{\l}a_{l}^{m}Y_{l}^{m}(\theta,\phi)
\]
with $a_{l}^{m\;\ast}\!=\!a_{l}^{-m}$. To satisfy Eqs. (\ref{eq:av_dis})
and (\ref{eq:corr_dis}), the real and imaginary parts of the
$a_{l}^{m}$ are drawn independently from a Gaussian distribution with 
variance $D=(dQ^{2}/k_{B}T)^{2}\Delta/2$ for $m\! \neq \! 0$  
and the $a_{l}^{0}$ from a Gaussian with variance $2D$.
Then the part of the Hamiltonian due to disorder can then be expressed 
\begin{equation}
\label{eq:disham}
\frac{{\cal H}_{dis}(\{v_{m}\})}{k_{B}T}= \sum_{p,q=0}^{N}\;
            \sum_{l=|p-q|}^{N}  
            a_{l,p-q}v_{p}^{\ast}v_{q}\;I_{l,(p+q-|p-q|)/2}^{|p-q|}
\end{equation}
where
\begin{eqnarray}
\label{eq:ilnm}
\lefteqn{I_{l,n}^{m}=\int d^{2}r Y_{l}^{m}\psi_{n+m}^{\ast}\psi_{n}}\\
&=&\!A_{n}\!A_{n\!+\!m}\sqrt{\frac{(2l\!+\!1)(l\!+\!m)!}{4\pi(l\!-\!m)!}}
             \frac{(-1)^{m}}{m!}B(N\!-\!n\!+\!1,n\!+\!m\!+\!1)\nonumber\\
&&\times\, _{3}F_{2}\left(\begin{array}{c}
             m\!-\!l,\, m\!+\!l+\!1,\,n\!+\!m+\!1,\nonumber\\
             m\!+\!1,\,N\!+\!m\!+\!2\,;\; 1 
             \end{array}\right)\nonumber
\end{eqnarray}
with $_{3}F_{2}$ a generalized hypergeometric function. This term of
the Hamiltonian limits our simulation in the case of disorder to a 
relatively small system size, as it requires finding and
storing $\sim  N^{3}/6$ values of  $_{3}F_{2}$.  
However, as the correlation length is very much reduced
by disorder, the effects of the finite system size are for the same
$\alpha_{T}$ much reduced compared with the non-disordered case. 

The time dependent Ginzburg-Landau equation, 
discretized in time, which drives our simulation is now:
\begin{eqnarray}
\lefteqn{v_{m}(t+\Delta t)-v_{m}(t)}\\ 
        &=&-\Delta t\,\Gamma \frac{\partial({\cal H}_{cl}(t)+
{\cal H}_{dis}(t))}
          {\partial v_{m}^{\ast}}
        +\Delta t\,\xi_{m}(t) \nonumber
\end{eqnarray}
with ${\cal H}_{cl}(t)$ and ${\cal H}_{dis}(t)$ 
as defined in Eqs. 
(\ref{eq:simham}) and (\ref{eq:disham}).

This equation
ensures that thermal averages can be replaced by time averages over
successive measurements from the simulation.
It leads to the correct canonical distribution after infinite time
in the limit $\Delta t\;\rightarrow\;0$ if real and imaginary parts of
the $\xi_{m}$ are drawn independently from a Gaussian probability 
distribution with variance $\sigma^2/\Delta t$ with
$\sigma^2=2\,\Gamma\,k_{B}T$ \cite{binney}. However, the finite
time steps always lead to a certain small deviation from the 
correct distribution, which does not arise in a Monte Carlo algorithm
\cite{moore}. We have tested our simulation's static thermal averages for 
varying 
quantities like energy and entropy against Monte Carlo averages from
the code used in Ref. \cite{dodgson}, and  chose 
our time steps $\Delta t=0.15$, which reduces deviations to less than 1.5\%.

The quenched disorder is 
self-averaging for the case of an infinite system.
 In a finite system though, the particular random choice of
disorder will obviously influence the results. To reduce this effect
we run the simulation  between 10 and 20 times
with different sets of random $a_{l}^{m}$, and average over these different
 measurements.

\section{THE DENSITY-DENSITY CORRELATOR}
\label{sec:dc}
To investigate the extent of order and the time scales of fluctuations in 
the system, we compute the connected part of the density-density 
correlator. This correlator  measures the correlations of the magnitude 
of the order parameter in space and time. The density-density correlator 
carries essentially the same information as the translational correlation 
function of the vortex positions \cite{o'neill}. 
However, it  is far easier to compute than the latter, because it does
not involve finding the zeros of the order parameter (see Section 
\ref{sec:sdc}). The density-density correlator is defined as
 
\begin{eqnarray}
\label{eq:strf}
S({\bf r'},t')&=&\langle|\psi({\bf r},t)|^{2}
                       |\psi({\bf r}+{\bf r'},t+t')|^{2}\rangle\\
              &&-\langle|\psi({\bf r},t)|^{2}\rangle
               \langle|\psi({\bf r}+{\bf r'},t+t')|^{2}\rangle \nonumber
\end{eqnarray} 
This correlator and its decay in time is most revealing in ${\bf k}$ space 
if
one is interested in the structure of the system. Therefore its 
Fourier transform is measured,
 normalized by the average density of the order parameter 
and for easier readability divided by its high temperature limit: 
\begin{equation}
\label{eq:Cdef}
C({\bf k}, t)=\frac{S({\bf k}, t)}{\langle|\psi|^{2}\rangle^{2}}\times
               \lim_{\alpha_{T}\rightarrow \infty}
                  \frac{\langle|\psi|^{2}\rangle^{2}}
                {S({\bf k},0)}
\end{equation} 

To compute this quantity the concept of the Fourier
transform on a plane has to be adapted 
to the curved two dimensional space of the
surface of a sphere.
The Fourier transform is an expansion of the correlation
function in the complete orthonormal set of functions solving the free wave 
equation, which is in the plane the continuous set of functions 
$e^{i{\bf k}\cdot {\bf r}}$.  The equivalent set of functions 
in a spherical geometry  is the discrete set of normalized spherical 
harmonics, $Y_{l}^{m}({\bf r})$. 
To a value of $l$ corresponds $k\!=\!l/R$. Because 
the system is isotropic, the correlator in {\bf k} space 
depends only on the magnitude of $k$, i.e. only on $l$ and not on $m$.
So, for better averaging we always calculate the correlator for all $m$
and average over the different $m$.  

\begin{equation}
\label{eq:strfofk}
S(l/R,t)=\frac{1}{2l+1}\sum_{m=-l}^{l}
            \int d^{2}r\,Y_{l}^{m}({\bf r})S({\bf r},t).
\end{equation} 

Now we  substitute for $S(l/R,t)$ from Eq. (\ref{eq:strf}), expand $\psi$  
in LLL eigenfunctions and take the spatial integral inside the
thermal average and the summation over the lowest Landau levels. 
The high temperature limit can easily be worked out analytically
\cite{dodgson} and the correlation function in Eq. \ref{eq:Cdef}
can be written in terms of thermal averages of the coefficients:  

\begin{figure}
\centerline{\epsfxsize= 8.5cm\epsfbox{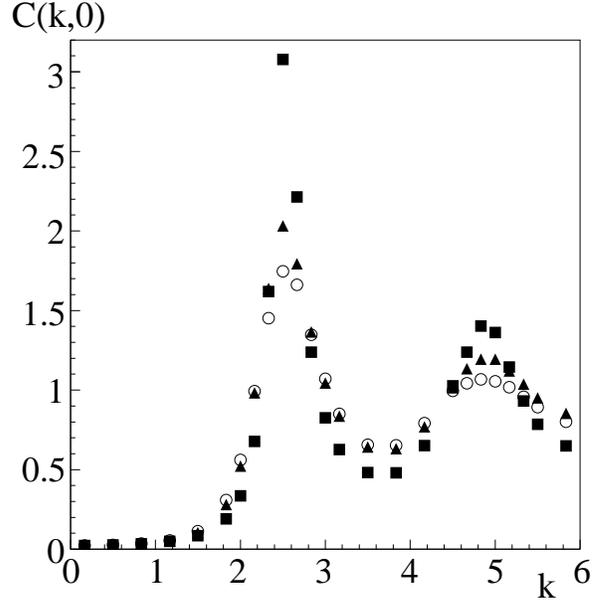}}
  \caption{Structure factor for $\alpha_{T}=-8$, 
     squares represent D=0 (pure limit),
    triangles  D=9, and circles D=25.}
  \label{figure1}
\end{figure}

\begin{eqnarray}
\label{eq:Cofv}
C(l/R, t\!-\!t')&=& \frac{4\pi(N-l)!(N+l+1)!}{(N!)^{2}(2l+1)}\,
                \frac{1}{ \langle\sum_{n=0}^{N}v_{n}^{\ast}v_{n}\rangle}
                \nonumber\\
            &\times&\!\sum_{m=-l}^{l}\left\langle
              \sum_{n=\max(0,-m)}^{\min(N,N-m)} 
        \!\!\!\!\!   v_{n+m}^{\ast}(t)v_{n}(t)I_{l,n}^{m}\right.\nonumber\\
            &&\times \!\!\!\left.\sum_{n'=\max(0,-m)}^{\min(N,N-m)}
        \!\!\!\!\!    v_{n'+m}^{\ast}(t')v_{n'}(t')I_{l,n}^{m}
              \right\rangle _{c},
\end{eqnarray} 
where $I_{l,n}^{m}$ is  defined in Eq. \ref{eq:ilnm}.
Here $\langle...\rangle_{c}$ signifies the connected part 
of the correlator, 
i.e.
$
\langle v_{i}^{\ast}(t)v_{j}^{\ast}(t')v_{k}(t)v_{l}(t')\rangle_{c}=
\langle v_{i}^{\ast}(t)v_{l}(t')\rangle
\langle v_{j}^{\ast}(t')v_{k}(t)\rangle .$
Without disorder
the  non-connected part of the density-density correlator,
 denoted $C_{nc}(k,t)$,
is zero for all $l\!\neq\!0$, which is equivalent to translational 
invariance, $\langle|\psi({\bf r},t)|^{2}\rangle$ constant.
However, in the disordered case there are permanent  
correlations due to the structure imposed on the vortices by the
quenched disorder. This is intuitively obvious: 
a permanent local potential will favor specific vortex constellations 
and thus result in permanent correlations.
 Therefore $\langle|\psi({\bf r},t)|^{2}\rangle$ is 
a function of ${\bf r}$ and $C_{nc}(k,t)$ is not zero.
The infinite time limit of the full correlator is equal to 
the non-connected part, for $t'\!\rightarrow\!\infty$
$
\langle|\psi({\bf r},t)|^{2}\psi({\bf r}+{\bf r'},t+t')|^{2}\rangle=
\langle|\psi({\bf r},t)|^{2}\rangle\langle|\psi({\bf r'},t')|^{2}\rangle.
$
We measure $C(k,t)\!+\!C_{nc}(k,t)$ directly. To
retrieve the connected part, $C_{nc}(k,t)$ is subtracted.

\subsection{The structure factor}
The density-density correlator $C(k,t)$ for $t'\!=\!0$ is the structure 
factor of the system. For the case of no disorder it has been 
investigated in detail by Dodgson and Moore \cite{dodgson}.
We give a short summary of their results, which we will extend by 
adding disorder.
The structure factor shows peaks (which sharpen 
as the temperature is lowered) near the inverse lattice
vectors of the triangular vortex lattice.
A Lorentzian always gives an excellent fit
 for the peak near the first
reciprocal lattice vector. The associated 
correlation length $\xi_{D}$, (D for density), is proportional 
to the inverse 
width of this peak at half its maximum, (denoted by $\delta^{-1}$),
and found to vary as $\xi_{D} \propto |\alpha_{T}|l_{m}$. 

Fig. \ref{figure1} shows the structure factor at the same effective
temperature for no, medium and very strong disorder. Disorder flattens
the peaks of the structure factor and 
makes it look rather like that of a 
clean system at a higher temperature. Like for the clean case we can fit 
the first peak to a Lorentzian. However, the region around the peak where
we get a good fit is narrower than in the clean case. 
Due to this limited fitting regime, 
the limited system size and, especially for strong disorder, insufficient
averaging over disorder, the errors are rather large. 
\begin{figure}
\centerline{\epsfxsize=9cm\epsfbox{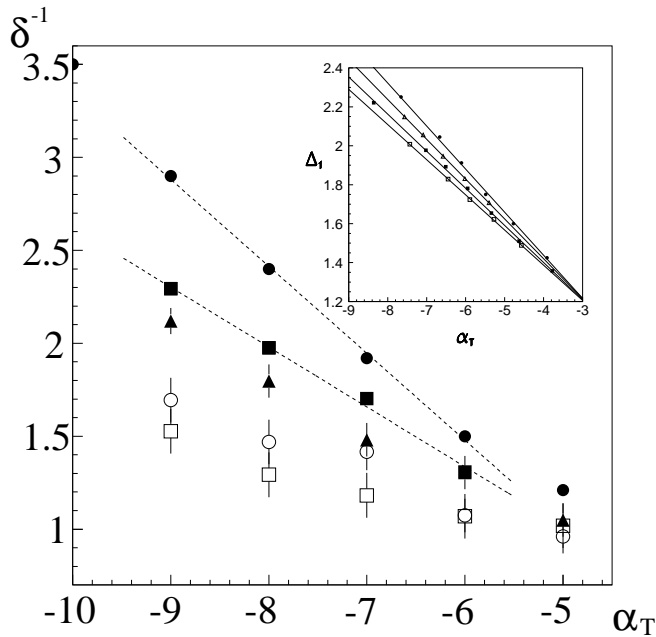}}
  \caption{The inverse width of the first peak in the structure factor 
    depending on $-\alpha_{T}$ with linear fits to the region 
    $-9 \leq\alpha_{T}\leq -6$.
    Filled circles, squares, triangles, empty circles 
    and squares
    correspond to D=0 (pure limit), D=0.25, 1, 4 and 9 respectively.
    For D=0 the system size is N=200, for D$\neq$0, N=72.
    The inset is reproduced from reference \protect \cite{Yeo}.
    It shows equivalent results of the parquet graph approximation 
    with linear fits for D=0, 0.1, 0.2 and 0.3. 
    }
  \label{figure2}
\end{figure}

Fig.\ref{figure2} shows the inverse peak width at half maximum, 
$\delta^{-1}\propto\xi_{D}$,
from these fits for different degrees of disorder. At high temperatures the
presence of disorder does not affect the correlation length $\xi_{D}$ much.
With decreasing temperature the growth of the correlation length 
is reduced. This effect is intuitively obvious because any hexagonal 
order is now opposed not only by thermal fluctuations but also by
the pinning of vortices to randomly distributed potential minima. 
A linear growth of $\delta^{-1}$ with $\alpha_{T}$ and a decrease 
of the gradient $|\partial(\delta^{-1})/\partial \alpha_{T}|$
with increasing disorder has been predicted using a non-perturbative,
parquet graph approximation to the two dimensional LLL 
system \cite{Yeo}. In the inset of Fig. \ref{figure2} a plot of 
$\delta^{-1}(\alpha_{T})$ from Ref. \cite{Yeo} is reproduced. 
We can  now compare
$\partial(\delta^{-1})/\partial \alpha_{T}$ from  the parquet graph 
approximation with our simulation results.
The absolute values of the gradients in the simulation are roughly 
a factor 2 larger than in Ref. \cite{Yeo}. 
However, the relative decrease of the gradient due to 
disorder agree well for Yeo and Moore's and
our results. If we refer to the linear fits in Fig. \ref{figure2}
for the case of an increase of disorder from D=0 to D=0.5,  
the change in our simulation is only a factor 1.15 larger. 

\begin{figure}
\centerline{\epsfxsize= 8.5cm\epsfbox{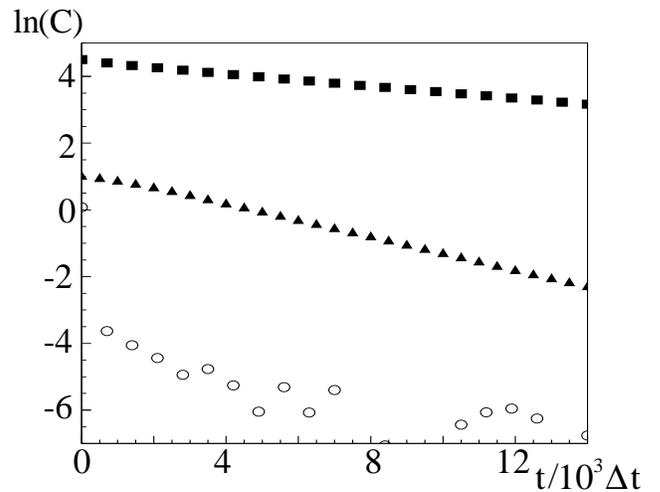}}
  \caption{Typical relaxation behavior for small k (k=0.5, circles) 
    and for larger k (k=2.5) 
    without disorder(squares) and with disorder (triangles)}
  \label{figure3}
\end{figure}

\subsection{Relaxation time scales}
To discuss the relaxation time scales of the density-density correlator
and their dependence on the wave vector, the effective temperature 
and the strength of disorder, 
we need to define a relaxation time $\tau_{D} (k,\alpha_{T}, D)$. 
Examples of the relaxation behavior of $C(k, t)$ with 
and without disorder 
are shown in Fig. \ref{figure3}. Without disorder
$\ln(C(k, t))$ shows an almost perfectly linear decay with time.
Only for very small times is the decay  faster than in a linear fit. 
In the case of disorder however, the decay of $\ln(C(k, t))$
is linear only for later times and has a smaller gradient at earlier times.
In this case we define $\tau_{D}$ from a fit only to the regime of truly 
linear exponential decay. We fit according to 
$C(k, t)\propto\exp(-t/\tau_{D})$ to extract the relaxation 
time scale $\tau_{D}$. This can be done with great accuracy for larger k.
However, with and without disorder, the error in this fit grows large 
for small k.
For $k\leq 1.5$ we have   $C(k,t)\!\ll \!1$. The relaxation 
times are rather small in this regime 
and $C(k,t)$ decays very quickly to zero. 
Very near zero, however, noise dominates the data, so that
in the worst cases we have to restrict our fits to the first
two or three points of the curve.   

Having defined $\tau_{D}$ from the exponential 
decay of the density-density 
correlator,
it is an interesting observation that the time scales which arise from
the exponential decay of the current correlator 
$\langle \nabla \! \times \! {\bf j}({\bf r},t) 
\nabla\! \times \! {\bf j}({\bf r'},t')\rangle$ 
which we measured when determining
the transverse conductivity \cite{transcond}
 are exactly the same. This suggests  
these time scales are important for all the dynamical
 properties of the system,
as is bourne out by our results on the self-diffusion coefficient
in Section \ref{sec:sdc}.

\begin{figure}
\centerline{\epsfxsize= 9cm\epsfbox{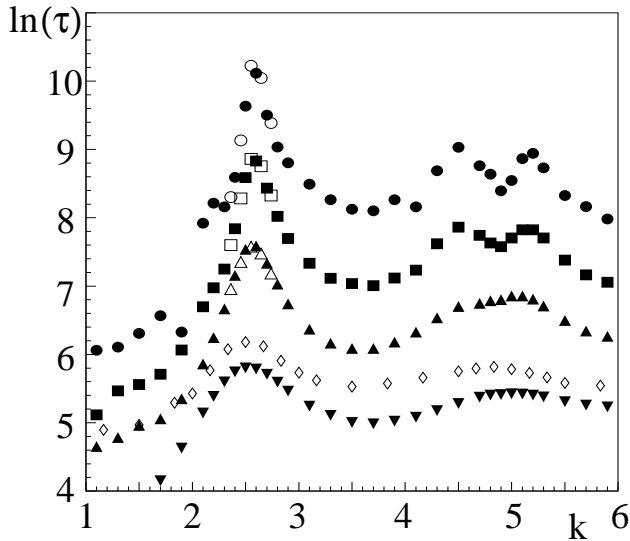}}
  \caption{Logarithmic Relaxation time scales 
    of the decay of the density-density correlator  
    in reciprocal space. Circles, squares,  
    triangles pointing up and down represent   D=0 (pure limit)  
    and $\alpha_{T} =$ -12,-10,-8 and -5 respectively.  
    For filled symbols system size 
    N=200, for empty symbols N=224.
    Diamonds represent D=4, $\alpha_{T} =$-8 and N=72.}
  \label{figure4}
\end{figure}

\begin{figure}
\centerline{\epsfxsize= 9cm\epsfbox{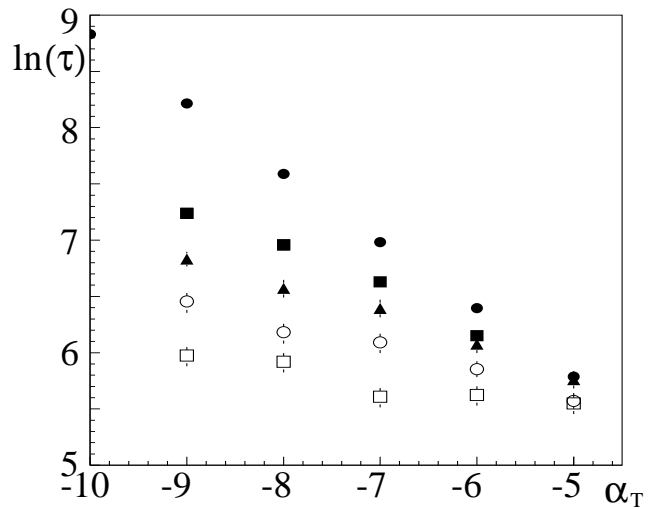}}
  \caption{Logarithm of the largest relaxation time scale versus $-\alpha_{T}$.
    Filled circles, filled squares, triangles, empty circles, and empty 
    squares
    correspond to D=0, 0.25, 1, 4, and 9 respectively. 
    For D=0 (pure limit) the system size is N=200, for D$\neq$0, N=72.}
  \label{figure5}
\end{figure}

We find that, in the vortex liquid regime which we are investigating,
$\tau_{D} (k,\alpha_{T}, D)$ reflects, like the structure factor, the
hexagonal lattice structure of the ground state. Fig. \ref{figure4}
shows a logarithmic plot of 
the k dependence of the relaxation times for different temperatures
and degrees of disorder.  Peaks can clearly be seen 
near the first, second and third reciprocal lattice vector 
of a triangular lattice in k space, which lie at k values of 2.694, 
4.665 and 5.387 \cite{o'neill}.
Note that for  the clean system at $\alpha_{T}\leq -10$ the second and 
third peak are clearly resolved, which is not the case for the structure 
factor in the same temperature regime.
For all measured temperatures, the longest relaxation time occurs at 
the first peak, i.e. $k\approx G=2.694$.  
If we introduce disorder to the system, the peaks near the reciprocal 
lattice
vectors flatten and, like the structure factor, the relaxation times look 
very similar to the ones in a clean sample at higher temperature. 
The error in $\tau_{D}$ increases with disorder for all values of k. 
Like in the case of the structure factor, this arises from insufficient 
averaging over disorder.     
The $\alpha_{T}$ dependence of the longest time
time scales $\tau_{D}(k\approx G)$ can be seen in a logarithmic plot
in Fig. \ref{figure5} for different degrees of disorder.
For the case without disorder the same quantity is also plotted in 
Fig. \ref{figure7} over the whole temperature range for which 
measurements were taken.

Without disorder we see 
an almost perfectly linear exponential increase with 
$|\alpha_{T}|$, i.e. $\tau_{D}(G) \propto \exp(C|\alpha_{T}|)$.
This indicates an activated relaxation process with an activation energy
$E_{a}/k_{B}T \propto \ln(\tau_{D})\propto|\alpha_{T}|$. 
With disorder a general decrease of the relaxation times 
can be observed. The effect of disorder is very small at high 
temperatures but grows stronger with decreasing $\alpha_{T}$. 
With weak disorder the relation $\ln(\tau_{D})\propto|\alpha_{T}|$
is, at least approximately, still valid, and  for strong disorder 
holds also within the range of the increasingly large error. 
Further discussion of these results is left to  Section
\ref{sec:ad}.

\section{SELF-DIFFUSION}
\label{sec:sdc}
We have measured the self-diffusion coefficient of the vortex positions 
${\bf R}$. In a liquid the square of the distance between 
initial position of a vortex  and its position after time t, 
denoted by S(t), increases linearly in 
time because of self-diffusion. The self-diffusion coefficient 
$D_{s}$ is then defined from the following relation:
\begin{equation}
\label{eq:selfdiff}
S(t)=\langle ({\bf R}(t)-{\bf R}(0))^{2}\rangle=2D_{s}\, t.
\end{equation}

In order to compute the self-diffusion coefficient we have to monitor 
the vortex
positions on the sphere of over not too short a time interval.
How the vortex positions depend on the LLL
coefficients can easily be seen if we make a coordinate transformation in 
Eq. (\ref{eq:phiexp}). We use the stereoscopic projection of the sphere 
into the complex plane, given by 
$z=x+iy=\cos(\phi) \tan(\theta /2)+i\sin(\phi) \tan(\theta /2)$. 
Now the order parameter expansion in Eq. (\ref{eq:phiexp}) 
can be written as 
\begin{equation}
\psi(z)= \frac{Q}{(2 \pi N)^{1/2}(|z|^{2}+1)^{N}} 
                      \sum_{m=0}^{N}v_{m}A_{m}z^{m}.
\end{equation}
The vortices are where $\psi(z)=0$, i.e. the roots of the
Nth order polynomial with coefficients $v_{m}A_{m}$. For a typical 
system size of N=200, the 100th coefficient in these polynomials 
is about 60 orders of magnitude larger than the first and the last, 
which makes the numerical root finding  nontrivial. With standard 
routines to find zeros of polynomials failing, we succeeded by 
searching for zeros on a very dense set of trial points using a Laguerre 
algorithm (see e.g. Ref. \cite{nrec}). 
However, we cannot always find all vortices on the sphere.   
Vortices very close to the south pole, 
which is  projected to infinity by the stereoscopic projection, 
can be inaccessible for the numerical routine.
The number of these inaccessible vortices increases  with system size.
With our method, for system sizes up to N=250 and for the nearly uniform
distribution  of vortices 
typically found, the maximum number of inaccessible roots 
is one, and for that one we know that it is very near the south pole. 

\begin{figure}
\centerline{\epsfxsize= 8.5 cm\epsfbox{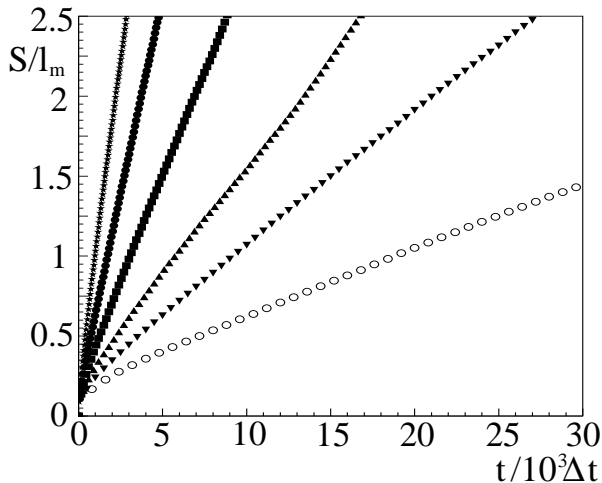}}
  \caption{$\langle ({\bf R}(t)-{\bf R}(0))^{2}\rangle$ against t
    for $\alpha_{T}=-7$ (stars), $\alpha_{T}=-8$ (dots), 
    $\alpha_{T}=-9$ (squares), 
    $\alpha_{T}=-10$ (triangles pointing up), 
    $\alpha_{T}=-11$ (triangles pointing down), 
    $\alpha_{T}=-12$ (circles)}
  \label{figure6}
\end{figure}

Once the roots are found, we have to identify the same vortex 
at different times in order to compute its self-diffusion. 
This was done by simply recording the vortex  positions in 
short time intervals $\Delta t$ and making a one-to-one mapping of the
positions at $t$ into the set at $t+\Delta t$ by pairing the vortices 
which are the smallest distance apart. 
Only very rarely these mappings were not one-to-one, but they could 
aways be made so by  pairing one new position with 
its second or third nearest old position.

\begin{figure}
\centerline{\epsfxsize= 9cm\epsfbox{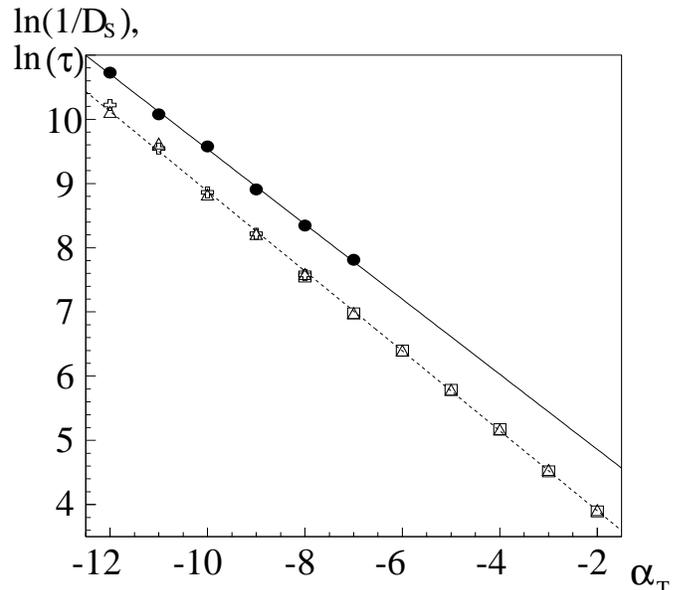}}
  \caption{Dominant time scales versus $-\alpha_{T}$ for D=0 
    (pure limit). 
    Empty squares, triangles and crosses:
    Logarithm of the largest relaxation time scale for system sizes 
    of N=72 ($2\leq -\alpha_{T}\leq 8$), 200 ($2\leq -\alpha_{T}\leq 12$),
    and 224 ($8\leq -\alpha_{T}\leq 12$) respectively.
    The dotted line is a linear fit for N=200.
    Filled circles:
    Logarithm of the inverse self-diffusion coefficient for N=200 
    with linear fit (solid line)}
  \label{figure7}
\end{figure}

We have recorded the self-diffusion for a system of 200 vortices and effective 
temperatures $\alpha_{T}$=-7,..,-12 for 13 ($\alpha_{T}$=-7) to 22 
($\alpha_{T}$=-12) times the density correlation relaxation time $\tau_{D}$.
A plot of S(t) as defined in Eq. (\ref{eq:selfdiff}) can be seen 
in Fig. \ref{figure6}. The dependence is linear except for very  
small times t. We have extracted $D_{s}$ from fits to the linear 
regime according to Eq. (\ref{eq:selfdiff}). We are interested in how the
diffusion time scale $\tau_{S\!D} \propto 1/D_{s}$ relates to the 
relaxation time scale of the density correlations $\tau_{D}$.
Fig. \ref{figure7} shows both time scales 
plotted  logarithmically against $\alpha_{T}$. The data suggests
 the same linear exponential 
dependence $\tau \propto \exp(C|\alpha_{T}|)$
holds for both time scales. 
If we extract the gradient C from the linear fits in Fig. 
\ref{figure7} for the data for $D_{s}$ and for $\tau_{D}$ 
for the same system size, they agree within 7\%.  
This agreement suggests that self-diffusion as well as
relaxation processes in the system are determined by the same
 mechanism, with an  activation energy that grows linearly 
with $-\alpha_{T}$. 

\section{ACTIVATED DYNAMICS}
\label{sec:ad}
We now put together our results regarding length and time scales
in the thin film. Firstly, the length scale of translational 
order  
$\xi_{D} \propto \delta^{-1}$ 
has been measured \cite{dodgson} and calculated \cite{Yeo} in 
the pure limit from the structure factor. It increases as 
$\xi_{D}\propto -\alpha_{T}$ at low 
temperatures. We found that this is  at least approximately true 
in a disordered system, a result  in good agreement with 
analytical results \cite{Yeo}. 
Secondly, there is one dominant  time scale $\tau$ which 
determines relaxation and self-diffusion times. It can be described  
as $\tau \propto \exp(-C \alpha_{T})$. Again this behavior is
not qualitatively changed by disorder.
These two results together yield 
\begin{equation}
\label{eq:as}
\tau \propto \exp(F\,\xi_{D}/l_{m}).
\end{equation}

\begin{figure}
\centerline{\epsfxsize= 9cm\epsfbox{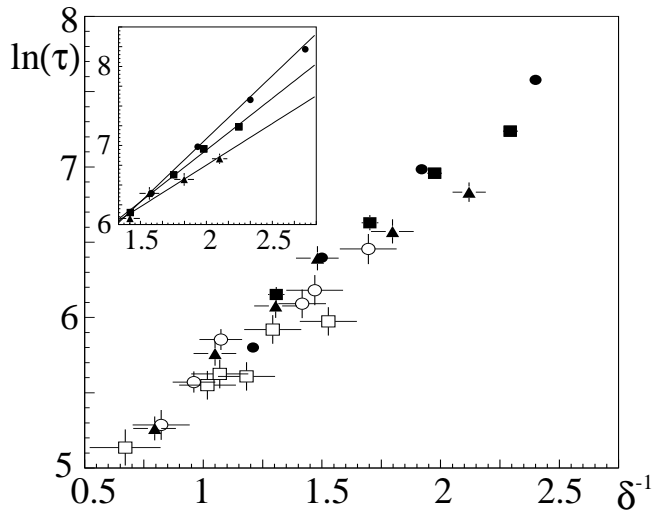}}
  \caption{The data from Figs. \protect \ref{figure5} and 
    \protect \ref{figure2} plotted as $\ln{\tau}$ versus $\delta^{-1}$.
    Filled circles, filled squares, triangles, empty circles, and empty 
    squares correspond to D=0, 0.25, 1, 4, and 9 respectively.
    The inset shows the data  for the temperature regime 
    $-9\leq \alpha_{T} \leq -6$ and weak disorder (D=0, 0.25, 1)
    with linear fits.}
  \label{figure8}
\end{figure}

Activated dynamics is said to occur when the time scale 
varies exponentially with some power 
 of the correlation length.
This kind of dynamics, which is found to be valid for example 
in spin glasses \cite{fisher&huse} is generally
described by  $\ln(\tau )\propto L^{\psi}$. Here L is the linear
 domain size, 
in our case given by $\xi_{D}$. The barrier-height
 exponent $\psi$
describes how the activation energy of the dominant relaxation process,
$E_{a}$, depends on the linear domain size. 
In our system $\psi$=1 and $E_{a} \propto k_{B}T \xi_{D}$. 
The proportionality constant $F$ depends on the strength of
disorder $D$. In Fig. \ref{figure8} we have plotted $\ln(\tau)$
versus $( \delta^{-1})$
for different disorder strengths. The poor quality of our data does 
only allow a quantitative description of $F(D)$. However, we can   
fit the data in the moderate to low temperature regime for weak disorder
according to Eq. (\ref{eq:as}) as shown in the inset of Fig. \ref{figure8}.
A clear  decrease of the gradient of the linear fits (and hence $F$) with
increasing disorder $D$ is visible.

\subsection{The relaxation mechanism}
\begin{figure}
\centerline{\epsfxsize= 8cm\epsfbox{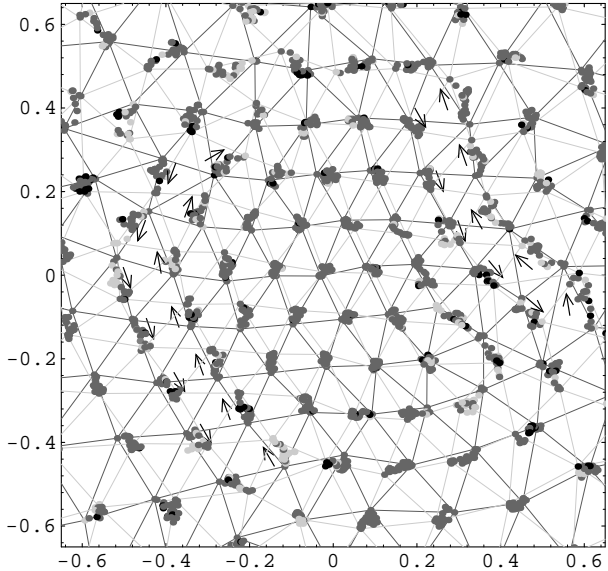}}
  \caption{Snap shots of the vortex motion in stereographic
    projection (r=0 is the north pole, r=1 is the equator)
    at $\alpha_{T}=-11$.
  The nearest-neighbor bonds are in darker grey for the starting
  points and in lighter grey for the end points. The vortex positions
  are shaded in lighter grey if their coordination number is 
  $\leq$5, in medium grey if it is 6 and in black if it is $\geq$7.
  Arrows point out opposite parallel motion of vortex chains . 
  } 
  \label{figure9}
\end{figure}

Fig. \ref{figure9} shows snap shots of the vortex dynamics at 
$\alpha_{T}=-11$. The vortex motion is shown over half a self-diffusion 
time $\tau_{S\!D}(\alpha_{T})$.    
After this time the average displacement of a vortex is $l_{m}$, which is 
$(\sqrt{3}/4\pi)^{1/2} \approx 0.37$ times the nearest
neighbor distance in the triangular lattice ground state. 
The vortex positions are shown for 24 time steps $\Delta t$
between initial and final position.
To cut out small random moves and thus make the overall displacement 
of a vortex more clearly visible, the positions in the picture at time t 
are an average of the vortex positions at times 
$t-\Delta t$, $t$ and $t+\Delta t$. The different grey shades 
of the vortex positions indicate their coordination number. 
The coordination numbers are not based on the time averaged, 
but on the real positions.
The nearest-neighbor bonds at initial and final positions are also shown.

To identify pairs of nearest neighbors and calculate the coordination
numbers we used a ``greedy'' 
triangulation algorithm. It uses the following definition of 
nearest neighbors in non-crystal context: 
Two vortices are nearest neighbors if their connection 
line does not cross with any other belonging to a pair a shorter 
distance apart. To apply this definition to a given configuration,
the vortices are paired in all possible combinations and all pairs  
ordered according to distance. Then it is decided successively, 
starting from the pair with the shortest distance, if the vortices of a 
pair qualify as nearest neighbors, in which case they are 
connected with a bond.  
A pair is not connected with a nearest neighbor bond
if its connection line crosses over one of the already existing 
nearest neighbor bonds. If this is not the case, the vortices 
in the pair are nearest neighbors and a bond is drawn.

We want to point out the main features about the vortex dynamics that can be
learned from  snap shots like the one in Fig. \ref{figure9}. 
At first sight they look quite  confusing, no easy patterns being 
distinguishable even after the vortex paths have been smoothed out. 
The real relaxation mechanisms are not nicely isolated processes in an 
otherwise crystalline environment. We do for example not see ``braids'' as 
suggested 
in reference \cite{dodgson}, which are formed by isolated motion of a 
ring of vortices. Where motion occurs, it does extend over a  region
of the plane. We also see that the vortices in the regions of 
motion change their coordination number very  often.
Topological  defects such as dislocations, equivalent to a pair of one 
fivefold and one sevenfold coordinated vortex, occur in large numbers
and their arrangements may be very transient, 
especially at higher temperatures. The kind of relaxation process we 
see takes place in a more or less strongly correlated liquid and 
is not well described in terms of  
isolated topological defects in an elastic environment. 

However, at a second look recurring themes in  the vortex motion 
become visible. The kind of structure which is easiest to spot is  
a chain-like motion of vortices. The process
underlying this chain motion can be identified as a 
tilt in the orientation of the nearest-neighbor grid 
in a strongly correlated region of the vortex liquid. 
In Fig. \ref{figure9} such a tilt is easiest to spot as a small clockwise
rotation about the mid point. However, a rotation of the   
underlying grid is not caused by a solid body rotation of all the
vortices in the region. For a large correlated region, a solid body rotation
would mean that vortices far from the center have to travel a 
long way, which does not happen. Instead each vortex moves   
in a way to reach a position in the new grid which is not too far 
from  its initial position. In Fig. \ref{figure9} this effect 
can be seen in the form of an opposite parallel 
motion of chains of vortices indicated with arrows. 

We are confident that we see 
a genuine property of the 2D liquid and
not a relaxation process favored by the fact that on the surface
of a sphere there is a permanent  presence of 12 
5-coordinated centers (for detailed discussion refer to \cite{dodgson}).
If these disclinations were crucial for the relaxation mechanism, one should
expect the relaxation times to be larger in larger systems, where 
these defects are less concentrated. Fig. \ref{figure7} however 
shows excellent agreement of $\tau_{D}$ for systems  different in size
by more than a factor 3. 

\subsection{Activation energy scaling}

We observe that the typical length of moving chains increases with  
decreasing $\alpha_{T}$. This is easy to understand because the
range of hexatic order $\xi_{D} \propto -\alpha_{T}$ determines the 
size of a tilt region.   
This pattern of fewer, large events of motion at low temperature 
and smaller, uncorrelated events at higher temperature is
confirmed by our self-diffusion measurements, where it can be 
deduced from the statistical 
error in the self-diffusion coefficient; 
if the simulation has been running for the same  time in units of 
relaxation times, $t/\tau(\alpha_{T})$=const., at different $\alpha_{T}$,
the vortices in the systems have diffused by the same amount but 
the error is clearly smaller at higher temperatures. 
The range of static correlations in the liquid determines
the size of a tilt region and therefore the energy barrier height 
of the relaxation mechanism. To explain why the activation energy  
$E_{a}$ depends linearly on $\xi_{D}$, 
we suggest the following scaling argument.

Consider a region of the vortex liquid with linear size $\xi_{D}$.
We want to estimate the energy it costs to tilt the core of 
the region, while the edges relax, as can be seen in Fig.\ref{figure9}, 
in order to keep best possible hexatic order with the static 
surroundings. Over a distance $\xi_{D}$ hexatic correlations 
from the middle of the region to the edges are only just noticeable.
Therefore the core of the region cannot rotate freely. The angles 
which allow some hexatic alignment with the surroundings 
are discrete in steps of $2\pi/N$, where $N$ is the number of vortices on 
the edge on the region. 
The potential energies of the $N$ different orientations 
with respect to the minimum energy orientation 
will have a probability distribution $P(E)$ with mean $\overline{E}$.
A region of size $\xi_{D}$ is typically in a low energy state just  
locked to a given orientation by its surroundings.
The locking energy, which is the lowest potential difference $E$ 
to one of the $N$ other orientations, denoted $E_{\min}$,
is of order $k_{B}T$. By rotating,
the system will jump from an initial low energy state 
to higher ones. 
From there, the rotated region and its 
surroundings relax to a new low energy  state, which is   
is the process responsible for destruction of correlations.  
The  intermediate ``barrier'' state after the rotation is 
one of the $N$ possible orientations, and its energy is  
a random energy from the distribution $P(E)$. We can write the  
typical activation energy  as $E_{a}=\overline{E}$.
From above we know that the typical lowest 
energy in a given sample $\overline{E_{\min}}\approx k_{B}T$.
In order to find $\overline{E}$ we can apply the well-known result 
that the minimum value of $E$ which occurs in a large 
random sample of size $N$ is on average  
$\overline{E_{\min}}\propto \overline{E}/N$.
This result is shown to be valid for any finite probability distribution  
$P(E)$, $E\geq 0$ with mean $\overline{E}$ in Ref. \cite{bray&moore}. 
We now know that $E_{a}=\overline{E}\propto N E_{\min}  \approx N k_{B}T$.  
$N$ is essentially the region's perimeter and therefore $N \propto \xi_{D}$,
which yields $E_{a}\propto \xi_{D}\,k_{B}T$, 
the scaling behavior observed in our measurements.

\section{CONCLUSIONS}
\label{sec:ccl}  

We have analyzed the equilibrium dynamics of a thin film superconductor 
in the vortex liquid phase.
We have shown that there is one main relaxation time scale which 
appears in the decay of density correlations of the order 
parameter as well as in the self-diffusion of the vortices.  
This time scale shows a linear exponential dependence on $\alpha_{T}$
at all accessible temperatures. 
We have also described the effects of quenched random disorder
on the density correlations and their decay. 
Disorder induces permanent correlations and reduces the range 
of the non-permanent correlations 
and their relaxation times. 
The density correlations in a disordered system  
look very similar to those in a system without disorder 
at a higher temperature. 
This is in good agreement with 
non-perturbative analytical results \cite{Yeo}.
We find that  both the pure and the disordered 
system have activated dynamics.
The main time scale $\tau$ scales with the range of
translational order $\xi_{D}$ like 
$\tau \propto \exp(F\, \xi_{D}/l_{m})$,
where F depends on the strength of disorder.
This indicates an activated relaxation process with an 
activation energy of the order
$E_{a}\propto k_{B}T\ \xi_{D}/l_{m}$. We have identified this
process as a tilt in the hexagonal orientation of the
vortex liquid over a region of linear extent $\xi_{D}$.
We have found no divergence of time scales or other features 
attributable to a phase transition at any finite temperature.  
Our work does not confirm the existence of a vortex lattice phase 
at finite temperature, nor do we see a vortex glass or Bragg glass phase 
in the disordered system.

Experimental results on time scales from ac transport measurement spectra 
in weakly disordered very films should provide information on
 the relaxation 
time scales found in our numerical work.
Such experiments have been performed, see for example 
Refs. \cite{yazdani} and \cite{kes_szef}.  
However, the samples in thin film experiments are are often two dimensional 
only in the sense that their characteristic bending length of vortices is 
larger than the sample thickness, but not thin enough to obey 2D LLL 
scaling. Therefore they do not compare directly to our results.  
Experiments similar to the ones from Refs.\cite{yazdani} and \cite{kes_szef}
but in thinner films would be directly related to our results.
Experimental observation of self-diffusion would be far more
difficult, if not impossible. Decoration experiments,
which use snap shots of the vortex liquid, have made 
detailed studies of static properties possible (see e.g. \cite{marchevsky} 
\cite{kim}). However, the analysis of the vortex dynamics with 
this technique remains difficult \cite{marchevsky}.

\section*{Acknowledgements}
We would like to thank Tom Blum
and Matthew Dodgson for useful interactions. 
AKK acknowledges financial support from a 
Manchester University 
Research Studentship and EPSRC.

\end{multicols}

\end{document}